\documentclass[twocolumn,superscriptaddress]{revtex4}
\usepackage{graphicx}
\usepackage{epstopdf}
\usepackage{amsmath}
\usepackage{booktabs}
\usepackage{color}
\usepackage{multirow}
\begin{document}

\title{Role of bond orientational order in the crystallization of hard spheres}

\author{John Russo} 
\affiliation{ {Institute of Industrial Science, University of Tokyo, Meguro-ku, Tokyo 153-8505, Japan} }

\author{Hajime Tanaka}
\affiliation{ {Institute of Industrial Science, University of Tokyo, Meguro-ku, Tokyo 153-8505, Japan} }

\begin{abstract}
With computer simulations of the hard sphere model, we examine in detail the microscopic pathway
connecting the metastable melt to the emergence of crystalline clusters. In particular we will show
that the nucleation of the solid phase does not follow a two-step mechanism, where crystals form inside dense precursor regions. On the contrary,
we will show that nucleation is driven by fluctuations of orientational order, and not by the density fluctuations.
By considering the development of the pair-excess entropy inside crystalline nuclei, we confirm that
orientational order precedes positional order.
These results are at odd with the idea of a two-step nucleation mechanism
for fluids without a metastable liquid-liquid phase separation. 
Our study suggests the pivotal role of bond orientational ordering in triggering crystal nucleation.
\end{abstract}

\maketitle

\section{Introduction}

After almost 150 years from the pioneering works of Gibbs, a satisfactory understanding of the
crystallization process is still lacking. Classical Nucleation Theory (CNT) describes nucleation as
an activated process in which the free energy gain of the solid phase over the fluid one
is in competition with the free energy penalty of forming an interface for nuclei of size smaller
than a characteristic length, the critical nucleus size. The two main limitations of CNT have
been recognized as the following: i) the \emph{capillarity approximation}, i.e. the assumption
that the thermodynamic properties of the small crystal nuclei are the same as the bulk crystal,
such as specific volume or surface tension;
ii) all the order parameters involved in the transition proceed simultaneously, and the transition
can be described effectively by one order parameter (as density for the nucleation
of liquid droplets from a supersaturated gas).
Many efforts have been done in order to relax some of the assumptions of CNT, showing in fact that
nuclei do not form at bulk condition and that the process is non-classical, i.e. the order parameters
involved in the transition do not change simultaneously~\cite{oxtoby}.

On these grounds, the two-step nucleation
mechanism has emerged as a candidate description for many of the discrepancies between CNT and experiments.
The idea is the following. The transition from the fluid phase to the solid phase involves at least
two order parameters, one associated with the breaking of translational symmetry and the other with
the breaking of rotational symmetry. In the crystallization of spherical particles, usually these
order parameters are identified with density and bond orientational order. In the two-step mechanism
scenario, the first fluctuation to trigger the transition is density fluctuation, which leads to the formation of a dense 
droplet in the melt. Then the droplet starts to be structured and a crystal
is formed. Large-amplitude density fluctuations can emerge from the criticality of a nearby liquid-liquid (L-L) phase
transition, which is the case of globular proteins~\cite{vekilov} or colloids with small range attractions~\cite{Savage}. This also
led to the speculation that critical fluctuations could enhance the crystallization rate~\cite{tenvolde,franzese}.
Rather surprisingly, the same mechanism was claimed for the nucleation pathway of many systems
for which a dense liquid phase is not known~\cite{myerson,pouget} or even does not exist at all~\cite{schope,schilling}.

Recently we have provided a detailed microscopic study of the nucleation pathway in ultrasoft potential systems~\cite{russo_gcm}
and in hard sphere systems~\cite{russo_hs}, both having purely repulsive potentials and thus without a L-L transition.
These works highlighted the role played by bond orientational order in the crystallization process~\cite{kawasaki}, finding no
evidence of the dense precursors upon which the two-step assumption is built. On the contrary, the transition resembles a
two-step process where the first step is the formation of extended structured regions of high orientational order, which
progressively densify. The bulk density is in fact reached only at a very late stage,
when the nucleus becomes several times the critical nucleus size. The first step, structuring at (almost) constant density
is achieved due to the wetting of bond-orientational regions to small crystalline nuclei. The identification
of the bond orientational order as the appropriate coordinate to describe the nucleation stage has unveiled
that the polymorph selection stage starts already in the metastable fluid phase before nucleation occurs~\cite{russo_gcm,russo_hs}.
With bond orientational order we have also studied the competition between crystalline structures and icosahedral
(or five-fold symmetric) structures~\cite{russo_hs,mathieu_russo}, i.e. the mechanism by which crystallization is prevented
and that paves the way to out-of-equilibrium states (i.e. glasses).

In the present contribution we will focus on the nucleation pathway in monodisperse hard spheres. We will show that, while density fluctuations
have a very short lengthscale, bond orientational order has an extended lengthscale that increases with supersaturation. The nucleation
events are localized inside the regions of high bond orientational order. The process is thus non-classical, with orientational order
preceding translational order as the nuclei grow.

\section{Methods}\label{sec:methods}
We run Umbrella Sampling Monte Carlo simulations of $N=4000$ monodisperse hard spheres
in the isothermal-isobaric (NpT) ensemble.
Hard spheres are ideal for studying crystallization and have already provided
tremendous contributions to our basic understanding of crystal nucleation~\cite{gasser,zaccarelli,schilling,sanz,valeriani,taffs}.
In the following, lengths are given in unit of the diameter $\sigma$, and
pressure in units of $k_{\rm B}T/\sigma^3$, where $k_{\rm B}$ is the Boltzmann constant.
We place the spheres randomly in a simulation box at packing fraction $\eta=0.5352$ and
equilibrate the system at reduced pressure $\beta p\sigma^3=17.0$. At this
pressure the liquid is metastable with respect to crystallization, with
a difference in chemical potential between the liquid and solid state of
$\beta |\Delta\mu|=0.54$~\cite{filion}. We run Umbrella Sampling simulations,
where a biasing potential is added to
the system Hamiltonian to sample crystalline clusters of different sizes, in order to
extract information on the free energy barrier and the critical cluster size.
Details on the implementation can be found in Ref.~\cite{AuerR}.
Configurations for clusters at different sizes are extracted from the simulations and analyzed.
The Umbrella Sampling biasing is used to gain better statistics by stabilizing nuclei at
the desired size. But we confirm that the same results are obtained if the configurations
are extracted from direct simulations (still possible at this pressure) without biasing potential.
At $\beta p\sigma^3=17$ the free energy barrier is $\beta\Delta F\simeq 18$
and the size of the critical nucleus is $n_c\simeq 80$.

The identification of crystalline particles follows the usual procedure~\cite{auer}.
A particle is identified as crystal if
its orientational order is coherent (in symmetry and in orientation) with that of its neighbors.
The $\ell$-fold symmetry of a neighborhood around each particle $i$ is characterized by a $(2\ell+1)$ dimensional complex vector ($\mathbf{q}_l$)
as $q_{\ell m}(i)=\frac{1}{N_b(i)}\sum_{j=1}^{N_b(i)} Y_{\ell m}(\mathbf{\hat{r}_{ij}})$, where
$\ell$ is a free integer parameter, and $m$ is an integer
that runs from $m=-\ell$ to $m=\ell$. The functions $Y_{\ell m}$ are the spherical harmonics
and $\mathbf{\hat{r}_{ij}}$ is the vector from particle $i$ to particle $j$.
The sum goes over all neighboring particles $N_b(i)$ of particle $i$. Usually 
$N_b(i)$ is defined by all particles within a cutoff distance, but in an inhomogeneous system
the cutoff distance would have to change according to the local density. Instead we 
fix $N_b(i)=12$ which is the number of nearest neighbors in icosahedra and close packed crystals (like \textsc{hcp} and \textsc{fcc})
which are known to be the only relevant crystalline structures for hard spheres.
The scalar product $(\mathbf{q}_6(i)/|\mathbf{q}_6(i)|)\cdot(\mathbf{q}_6(j)/|\mathbf{q}_6(j)|)$ quantifies the similarity of the two environments.
If it exceeds $0.7$ between
two neighbors, they are deemed \emph{connected}. We then identify a particle as crystalline if it is connected with at least $7$ neighbors~\cite{auer}.

\begin{figure}
 \centering
 \includegraphics[width=5cm]{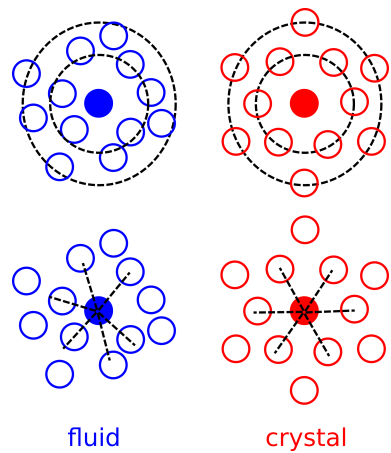}
 \caption{Translational order (upper row) and orientational order (bottom row) in a fictitious
 liquid (left column) and crystal (right column) configurations.
 Translational order expresses the relative spacing between particles in the system, while
 orientational order the relative orientation between the neighbors around each particle.}
 \label{fig:op}
\end{figure}

The liquid-to-solid transition is characterized by the symmetry breaking of both translational
and orientational order~\cite{torquato}. Translational order expresses the relative spacing between particles in the system,
as shown in the top row of Fig.~\ref{fig:op}. A good order parameter for translational order is the local density $\rho$,
which is calculated by constructing the Voronoi diagram and measuring
the volume of the cell associated with each particle. More generally, translational order can be obtained
from two-body correlation functions, such as the pair correlation function $g(r)$~\cite{torquato}. Since we are dealing with hard sphere
systems, where there is no energy involved, the most meaningful definition of translational order
comes from the two-body excess entropy, defined as:
\begin{equation}\label{eq:s2}
s_2=-\frac{\rho}{2}\int dr\left[g(r)\log(g(r))-g(r)+1\right]
\end{equation}

Orientational order, instead, expresses the relative orientation between the neighbors around each particle,
as shown in the bottom row of Fig.~\ref{fig:op}. Unlike translational order, which is obtained
from two-body correlation functions, orientational order is obtained by considering many-body correlations.
For hard spheres it is well known that the local bond-order analysis introduced by
Steinhardt~\cite{steinhardt} is an adequate measure of orientational order. This order parameter is obtained
by constructing a rotational invariant from the quantity $q_{\ell m}(i)$ previously introduced. In the present
work we are going to focus on the coarse-grained quantity $Q_6$, which can be obtained in the following way.
First $q_{\ell m}(i)$ is spatially coarse-grained~\cite{lechner}
\begin{equation}\label{eq:Qlm}
	Q_{\ell m}(i) = \frac{1}{N(i)+1}\left( q_{\ell m}(i) +  \sum_{j=0}^{N(i)} q_{\ell m}(j)\right), 
\end{equation}
where $N(i)$ are the neighbors of particle $i$. Then the following invariant is obtained
\begin{equation}\label{eq:ql}
	Q_\ell = \sqrt{\frac{4\pi}{2l+1} \sum_{m=-\ell}^{\ell} |Q_{\ell m}|^2 }, 
\end{equation}
where we focus on the case $l=6$ (corresponding to the six-fold symmetry of the crystal structures).
The effect of coarse graining $Q_6$ has proven effective not only in reducing the signal-to-noise
ratio~\cite{lechner}, but also eliminates the signal from five-fold symmetric structures~\cite{russo_hs,mathieu_icosahedra} which do not 
add coherently. In this way the order parameter goes continuously from low values in the fluid phase to high
values in the crystal phase.

\begin{figure}
 \centering
 \includegraphics[width=7.5cm]{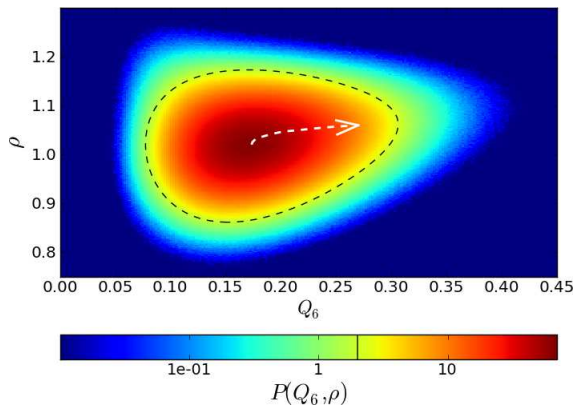}
 \caption{Probability distribution for the metastable phase in the $\rho-Q_6$ space. The dashed black line
 is a contour line. The dashed white arrow is instead a steepest descent path from the maximum to a high $Q_6$ point 
 of the probability distribution function.}
 \label{fig:map}
\end{figure}

\section{Results}\label{sec:results}

We start by looking at the two-dimensional probability distribution of density $\rho$ and bond orientational
order $Q_6$, for a metastable fluid state at pressure $\beta p\sigma^3=17$ (before the appearance of the
critical nucleus), reported in Fig.~\ref{fig:map}. The probability distribution is related to the Landau free
energy, $F(Q_6,\rho)=-k_{\rm B}T \log P(Q_6,\rho)$. The free energy can be well fitted with a
full cubic polynomial, for which the most important term is of the form $Q_6\rho^2$. This term is responsible
for the shape of contours lines (dashed line in Fig.~\ref{fig:map}): because the interaction is linear
in $Q_6$ and quadratic in $\rho$, the system can increase its orientational order without an increase of its
translational order, but the contrary is not true, and an increase in density also increases the average $Q_6$.
A similar term was proposed in order to describe quasi-crystals formation, where the ordering of the orientational
field does not imply an ordering of the translational field~\cite{jaric}. Note also that a small linear coupling
between $Q_6$ and $\rho$ exists at high $Q_6$, seen for example in the small slope of the steepest descent path
(white dashed arrow) in Fig.~\ref{fig:map}. As we will discuss later, we believe this small coupling term to be responsible
for previous observations of two-step nucleation processes in hard spheres.

\begin{figure}[!t]
 \centering
 \includegraphics[width=7.5cm]{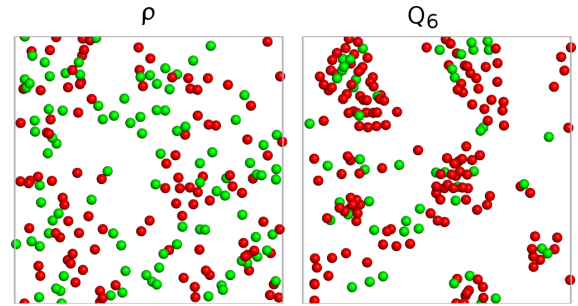}
 \caption{Same configuration but with different sets of particles plotted. (left) $5\%$ of particles having
 the highest value of $\rho$. Colors are given by sorting all the particles (also the ones not plotted) according to their value of $Q_6$, and dividing
 the set in two halves: red (dark gray) particles have a high value of $Q_6$, while green (light gray) particles a low value of $Q_6$.
 (right) $5\%$ of particles having
 the highest value of $Q_6$. Colors are given according to the value of $\rho$ of each particle: red (dark gray) particles
 have a high value of $\rho$, while green (light gray) particles have a low value of $\rho$.
 }
 \label{fig:lengths}
\end{figure}

Next we address the correlation length of these order parameters. In the two-step scenario, crystal nucleation
occurs inside dense droplets that form because of fluctuations in the density field. In a recent study we have
compared correlation lengths for both translational and orientational order~\cite{mathieu_russo} and found that,
while the correlation length associated with density (and any two-body quantity) is always very short and does not
grow with increasing supersaturation, the correlation length of orientational order is an increasing function
of supersaturation. This already suggests that the mechanism
of formation of dense droplets is unlikely in hard spheres, while a viable mechanism would be the nucleation inside
localized regions of high orientational order~\cite{russo_hs,kawasaki}.
We show this in Fig.~\ref{fig:lengths}, by simply looking at a typical configuration in the metastable state.
On the left panel we plot a subset of particles ($5\%$ of the total) having the highest value of $\rho$. It is
immediately clear that the position of particles is approximately uncorrelated, without any apparent lengthscale.
The colors are given according to the bond orientational order field, with high $Q_6$ particles being colored
in red (dark gray), and low $Q_6$ particles colored in green (light gray). We see that there is a lack of correlation
of particles in dense environments and the value of $Q_6$. Going back to Fig.~\ref{fig:map} this is confirmed by observing that, at
high density, the $Q_6$ value (while increasing on average) is equally likely to have low or high values.
On the right panel of Fig.~\ref{fig:lengths} we plot instead the subset of particles ($5\%$ of the total) having the highest value of $Q_6$.
Unlike the density field, bond orientational order clearly displays a strong correlation between the particles,
with structures resembling droplets. In this case, the colors are given according to the value of $\rho$ of each particle:
red (dark gray) particles having a high value of $\rho$ and green (light gray) particles a low value of $\rho$. This time there is a clear correlation
between high values of $Q_6$ and high values of $\rho$, so clusters of high orientational order will appear denser. This correlation
comes from the linear term that we noted in the distribution function of Fig.~\ref{fig:op}.
\begin{figure}
 \centering
 \includegraphics[width=7cm]{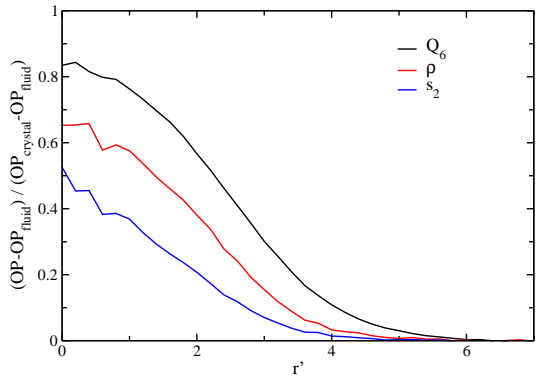}
 \caption{Radial profile of the averaged critical nucleus (nuclei with size between $80$ and $90$ particles). The profiles are
 for different order parameter (OP) fields (from top to bottom): $OP=Q_6$ (black line), $OP=\rho$ (red or dark gray line) 
 and $OP=s_2$ (blue or light gray line). The fields are normalized as
 to be $0$ in the fluid phase and $1$ in the bulk crystal phase. As usual, distances are in unit of the hard sphere diameter $\sigma$.}
 \label{fig:profile}
\end{figure}
As we will show soon, since
crystals are born from fluctuations of the orientational order field, a measure of the density of the regions from
which the crystal originates would give a value higher than the average melt density. But this does not mean that the
crystal originated due to a fluctuation of the density field, but it is simply a consequence of this small linear coupling between
$\rho$ and $Q_6$.

We now proceed to study the nucleation stage, where a crystal is nucleated up to the critical size, eventually overcoming the
free energy barrier and starting the crystallization process. We first focus on nuclei of a size comparable to the critical
nucleus size (so we average over many independent configurations where the biggest nucleus is of size between $80$ and $90$
particles). In Fig.~\ref{fig:profile} we plot the radial profile for three different fields, $Q_6$, $\rho$ and the two-body excess entropy
$s_2$, as a function of the distance from the center of the nucleus ($r'$).  The fields are normalized as to be $0$ in the fluid phase and $1$ in the bulk crystal phase. First we note that the process
is non-classical, where all fields have still not reached their bulk value, and the radial distribution is different for each field.
In particular, going from the fluid phase (high value of $r'$) to the center of the crystal nucleus ($r'=0$),
the $Q_6$ field is the first to develop, while density is lagging behind the development of the orientational field.
\begin{figure}
 \centering
 \includegraphics[width=7cm]{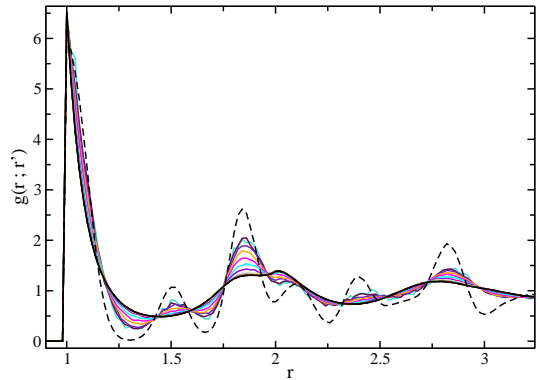}
 \caption{Pair correlation function $g(r;r')$ calculated at different distances $r'$ (in steps of $\Delta r'=0.5\sigma$) from the critical nucleus center.
 The $g(r,r')$ at the center of the critical nucleus ($r'=0$) is the one closest to the dashed line, which is
 the pair correlation function for a bulk $rhcp$ crystal. Distances are in unit of the hard sphere diameter $\sigma$.}
 \label{fig:gr}
\end{figure}
This again proves that nucleation is driven by orientational order and not by density. The same result is confirmed by looking at other definitions
of translational order, which can be extracted from two-body correlation functions. For hard spheres systems, where there is no energy
term in the free energy, a physically motivated definition of translational order is given by the two-body excess entropy $s_2$ defined
in Eq.~\ref{eq:s2}. To obtain the radial dependence of $s_2(r')$ we introduce the functions $g(r;r')$ which
measure the radial distribution function averaged over particles at distance $r'$ from the critical nucleus.
In formula
$$
\rho g(r;r')=\frac{1}{N}\langle{\sum_{r'<r_i<r'+dr'}^N\sum_{j\neq i}^N\delta (\mathbf{r}-\mathbf{r}_i+\mathbf{r}_j)}\rangle^{\ast}
$$
where the ensemble average $\langle\cdots\rangle^{\ast}$ is over Umbrella Sampling configurations with a crystal nucleus of critical size.
The two-body excess entropy function is then
$$
s_2(r')=-\frac{\rho}{2}\int dr\left[g(r;r')\log(g(r;r'))-g(r;r')+1\right]
$$
where the integral is carried up to a distance corresponding to the second nearest-neighbors shell (as nuclei are still small).
The result shown in Fig.~\ref{fig:profile} clearly indicates that the translational order (expressed by $s_2$) is lagging behind orientational
order (expressed by $Q_6$).

The slow positional ordering inside the critical nucleus can be seen in Fig.~\ref{fig:gr}, where the pair correlation function $g(r,r')$ is plotted
at different distances ($r'$) from the center of the critical nucleus. The $g(r,r')$ at the center is the one closest to the dashed line, which is the pair
correlation function $g(r)$ of a
bulk $rhcp$ crystal. Even at the center of the critical nucleus ($r'=0$), the pair correlation function is far from that of a bulk crystal.

\begin{figure}
 \centering
 \includegraphics[width=7cm]{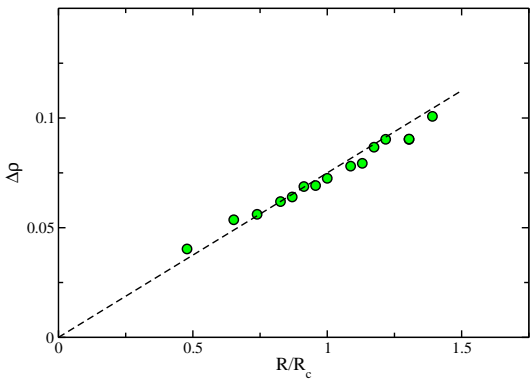}
 \caption{Excess density at the center of the nucleus ($\Delta\rho=\rho(r=0)-\rho_{fluid}$) 
 as a function of the size of the nucleus $R$, with $R_c=2.3\,\sigma$ being the critical
 nucleus radius. The increase in density from the fluid phase up to nuclei bigger than the critical size is linear, with $\Delta\rho(R=0)=0$.}
 \label{fig:rho_center}
\end{figure}

Already from Fig.~\ref{fig:profile} it is clear that the two-step scenario with dense precursors is unlikely, as orientational order precedes the
densification process. To have a direct confirmation of the absence of a two-step process, in Fig.~\ref{fig:rho_center} we plot the density at the center of the nucleus
as a function of the nucleus size. The increase in density is linear with the size of the nucleus, and extrapolates to the fluid density in the limit 
of vanishing size of
the nucleus. This result is in contrast to the expectation of the two-step nucleation mechanism, which instead predicts a non-linear dependency of the
density on the nucleus size: first an enrichment at constant size, followed by a size growth with little density change. Figure~\ref{fig:rho_center} thus confirms that
there is nothing anomalous with the density increase during the nucleation process.

\section{Conclusions}\label{sec:discussion}
In the present article we have examined in detail the possibility of a two-step nucleation process from the supercooled state in hard spheres.
The two-step nucleation is based on a specific behavior of two order parameters: density and structure. The fluid phase first forms a dense precursor,
which is then structured to form a crystalline nucleus. Nucleation is then initiated by density fluctuations, and followed by the superposition of
a structural fluctuation. Unlike this scenario, we stress that hard spheres take another route to crystal nucleation. 
Nucleation is promoted by extensive fluctuations
in bond orientational order. These fluctuations act as precursors for the formation of crystalline nuclei. The increase of density inside these regions lags behind
the development of orientational order, so that one can effectively have small crystalline nuclei with densities still very far from bulk conditions.
The reason why the transition from a fluid structure to a solid structure occurs with little density change is due to the fact that nuclei form in regions
of high orientational order, and so are wetted by fluid regions with a high value of this field. The structuring of the nuclei then requires just small adjustments
of particle positions, without big density changes \cite{russo_hs}.

This decoupling between the orientational order parameter and the translational order parameter has also important consequences for the glass transition. Once
the positional order transition is avoided, orientational order keeps growing with increasing supersaturation. It has been recently shown that this increase
of orientational order is linked with the slowing down of the dynamics, displaying critical-like behavior~\cite{tanaka,mathieu_icosahedra}.

Interestingly, the mechanism by which the positional ordering transition is avoided seems to be also correlated with the development of orientational order.
We have recently shown~\cite{russo_hs} that crystalline structures are not the only structures that are stabilized by an increase of orientational order.
There is in fact a competition between crystalline structures and icosahedral structures (five-fold symmetric structures), and we have shown that both an
increase in density and polydispersity~\cite{mathieu_russo} stabilize the icosahedral structure, effectively suppressing crystallization.
This again confirms the effect of frustration on crystallization \cite{TanakaGJPCM}, 
as in the case of 2D spin liquids~\cite{ShintaniNP,STNM}, and as recently observed
in metallic glasses~\cite{Jakse2008,Hwang2012}.

We stress that the two-step scenario is instead fully consistent for systems which display a L-L transition. In this case, the presence of a minima
(even if metastable) in the system's free energy corresponding to the liquid phase can start the nucleation
process without a macroscopic phase separation~\cite{tenvolde,vekilov,Savage,desgranges,murata}.


\section{Acknowledgments}
 This study was partly supported by a grant-in-aid from 
the Ministry of Education, Culture, Sports, Science and Technology, Japan (Kakenhi)
and by the Japan Society for the Promotion of
Science (JSPS) through its ``Funding Program for World-Leading
Innovative R\&D on Science and Technology (FIRST Program)'' and a JSPS Postdoctoral Fellowship.

%
%

\begin{thebibliography}{34}
\expandafter\ifx\csname natexlab\endcsname\relax\def\natexlab#1{#1}\fi
\providecommand{\enquote}[1]{``#1''}
\expandafter\ifx\csname url\endcsname\relax
  \def\url#1{\texttt{#1}}\fi
\expandafter\ifx\csname urlprefix\endcsname\relax\def\urlprefix{URL }\fi
\providecommand{\eprint}[2][]{\url{#2}}

\bibitem[Oxtoby(2003)]{oxtoby}
D.~Oxtoby, \emph{Philos. T. R. Soc. A} \textbf{361}, 419--428 (2003).

\bibitem[Galkin and Vekilov(2000)]{vekilov}
O.~Galkin, and P.~Vekilov, \emph{Proc. Nat. Acad. Sci. U.S.A.} \textbf{97},
  6277--6281 (2000).

\bibitem[Savage and Dinsmore(2009)]{Savage}
J.~R. Savage, and A.~D. Dinsmore, \emph{Phys. Rev. Lett.} \textbf{102}, 198302
  (2009).

\bibitem[ten Wolde and Frenkel(1997)]{tenvolde}
P.~ten Wolde, and D.~Frenkel, \emph{Science} \textbf{277}, 1975--1978 (1997).

\bibitem[Xu et~al.(2012)]{franzese}
L.~Xu, S.~V. Buldyrev, H.~E. Stanley, and G.~Franzese, \emph{Phys. Rev. Lett.}
  \textbf{109}, 095702 (2012).

\bibitem[Garetz et~al.(2002)]{myerson}
B.~A. Garetz, J.~Matic, and A.~S. Myerson, \emph{Phys. Rev. Lett.} \textbf{89},
  175501 (2002).

\bibitem[Pouget et~al.(2009)]{pouget}
E.~Pouget, P.~Bomans, J.~Goos, P.~Frederik, G.~de~With, and N.~Sommerdijk,
  \emph{Science} \textbf{323}, 1455--1458 (2009).

\bibitem[Sch{\"o}pe et~al.(2006)]{schope}
H.~Sch{\"o}pe, G.~Bryant, and W.~van Megen, \emph{Phys. Rev. Lett.}
  \textbf{96}, 175701 (2006).

\bibitem[Schilling et~al.(2010)]{schilling}
T.~Schilling, H.~J. Sch\"ope, M.~Oettel, G.~Opletal, and I.~Snook, \emph{Phys.
  Rev. Lett.} \textbf{105}, 025701 (2010).

\bibitem[Russo and Tanaka(2012{\natexlab{a}})]{russo_gcm}
J.~Russo, and H.~Tanaka, \emph{Soft Matter} \textbf{8}, 4206--4215
  (2012{\natexlab{a}}).

\bibitem[Russo and Tanaka(2012{\natexlab{b}})]{russo_hs}
J.~Russo, and H.~Tanaka, \emph{Scientific Reports} \textbf{2}
  (2012{\natexlab{b}}).

\bibitem[Kawasaki and Tanaka(2010)]{kawasaki}
T.~Kawasaki, and H.~Tanaka, \emph{Proc. Nat. Acad. Sci. U.S.A.} \textbf{107},
  14036 (2010), ISSN 0027-8424.

\bibitem[Leocmach et~al.(2012)]{mathieu_russo}
M.~Leocmach, J.~Russo, and H.~Tanaka, \emph{Importance of many-body
  correlations in glass transition: an example
	from polydisperse hard spheres. Accepted in J. Chem. Phys.}  (2012).

\bibitem[Gasser et~al.(2001)]{gasser}
U.~Gasser, E.~R. Weeks, A.~Schofield, P.~N. Pusey, and D.~A. Weitz,
  \emph{Science} \textbf{292}, 258 (2001).

\bibitem[Zaccarelli et~al.(2009)]{zaccarelli}
E.~Zaccarelli, C.~Valeriani, E.~Sanz, W.~C.~K. Poon, M.~E. Cates, and P.~N.
  Pusey, \emph{Phys. Rev. Lett.} \textbf{103}, 135704 (2009).

\bibitem[Sanz et~al.(2011)]{sanz}
E.~Sanz, C.~Valeriani, E.~Zaccarelli, W.~C.~K. Poon, P.~N. Pusey, and M.~E.
  Cates, \emph{Phys. Rev. Lett.} \textbf{106}, 215701 (2011).

\bibitem[Valeriani et~al.(2012)]{valeriani}
C.~Valeriani, E.~Sanz, P.~Pusey, W.~Poon, M.~Cates, and E.~Zaccarelli,
  \emph{Soft Matter} \textbf{8}, 4960 (2012).

\bibitem[Taffs et~al.(2012)]{taffs}
J.~Taffs, S.~Williams, H.~Tanaka, and C.~Royall, \emph{arXiv preprint
  arXiv:1206.5526}  (2012).

\bibitem[Filion et~al.(2010)]{filion}
L.~Filion, M.~Hermes, R.~Ni, and M.~Dijkstra, \emph{J. Chem. Phys.}
  \textbf{133}, 244115 (2010).

\bibitem[Auer and Frenkel(2005)]{AuerR}
S.~Auer, and D.~Frenkel, \emph{Adv. Polym. Sci.} \textbf{173}, 149--207 (2005).

\bibitem[Auer and Frenkel(2004)]{auer}
S.~Auer, and D.~Frenkel, \emph{J. Chem. Phys.} \textbf{120}, 3015--3029 (2004).

\bibitem[Truskett et~al.(2000)]{torquato}
T.~M. Truskett, S.~Torquato, and P.~G. Debenedetti, \emph{Phys. Rev. E}
  \textbf{62}, 993 (2000).

\bibitem[Steinhardt et~al.(1983)]{steinhardt}
P.~J. Steinhardt, D.~R. Nelson, and M.~Ronchetti, \emph{Phys. Rev. B}
  \textbf{28}, 784--805 (1983).

\bibitem[Lechner and Dellago(2008)]{lechner}
W.~Lechner, and C.~Dellago, \emph{J. Chem. Phys.} \textbf{129}, 114707 (2008).

\bibitem[Leocmach and Tanaka(2012)]{mathieu_icosahedra}
M.~Leocmach, and H.~Tanaka, \emph{Nature Commun.} \textbf{3}, 974 (2012), ISSN
  2041-1723.

\bibitem[Jari\ifmmode~\acute{c}\else \'{c}\fi{}(1985)]{jaric}
M.~V. Jari\ifmmode~\acute{c}\else \'{c}\fi{}, \emph{Phys. Rev. Lett.}
  \textbf{55}, 607--610 (1985).

\bibitem[Tanaka et~al.(2010)]{tanaka}
H.~Tanaka, T.~Kawasaki, H.~Shintani, and K.~Watanabe, \emph{Nature Mater.}
  \textbf{9}, 324--331 (2010).

\bibitem[Tanaka(1998)]{TanakaGJPCM}
H.~Tanaka, \emph{J. Phys.: Condens. Matter} \textbf{10}, L207--L214 (1998).

\bibitem[Shintani and Tanaka(2006)]{ShintaniNP}
H.~Shintani, and H.~Tanaka, \emph{Nature Phys.} \textbf{2}, 200--206 (2006).

\bibitem[Shintani and Tanaka(2008)]{STNM}
H.~Shintani, and H.~Tanaka, \emph{Nature Mater.} \textbf{7}, 870--877 (2008).

\bibitem[Jakse and Pasturel(2008)]{Jakse2008}
N.~Jakse, and A.~Pasturel, \emph{Phys. Rev. B} \textbf{78}, 214204 (2008).

\bibitem[Hwang et~al.(2012)]{Hwang2012}
J.~Hwang, Z.~Melgarejo, Y.~Kalay, I.~Kalay, M.~Kramer, D.~Stone, and P.~Voyles,
  \emph{Phys. Rev. Lett.} \textbf{108}, 195505 (2012), ISSN 0031-9007.

\bibitem[Desgranges and Delhommelle(2011)]{desgranges}
C.~Desgranges, and J.~Delhommelle, \emph{J. Am. Chem. Soc.} \textbf{133},
  2872--2874 (2011).

\bibitem[Murata and Tanaka(2012)]{murata}
K.~Murata, and H.~Tanaka, \emph{Nature Mater.} \textbf{11}, 436--443 (2012).

\end{thebibliography}

\end{document}